\documentclass[%
    reprint,
    twocolumn,
    aps,
    physrev,
    superscriptaddress,
]{revtex4-2}
\usepackage{graphicx}
\usepackage{microtype}              
\usepackage[utf8]{inputenc}         
\usepackage[T1]{fontenc}            
\usepackage{newtxtext}              
\usepackage[smallerops]{newtxmath}  
\usepackage{booktabs}
\usepackage{enumitem}
\usepackage[dvipsnames]{xcolor}
\usepackage{hyperref}
\hypersetup{
    colorlinks=true,
    linkcolor=NavyBlue,
    citecolor=NavyBlue,
    urlcolor=NavyBlue
}

\def\iid{iid}
\def\snr{\textsc{snr}}
\def\rhs{\emph{rhs}}

\newcommand*{\tran}{^{\mkern-1.5mu\mathsf{T}}}
\DeclareMathOperator{\tr}{Tr}

\DeclareMathOperator{\diag}{diag}
\DeclareMathOperator*{\E}{\mathbf{E}}
\DeclareMathOperator*{\Cov}{\mathrm{Cov}}

\newcommand{\equalcontrib}{\thanks{%
    These authors contribute equally to this work.%
    }%
}
\begin{document}\frenchspacing
\title{Data coarse graining can improve model performance}
\author{Alex Nguyen}
    \affiliation{%
    Princeton Neuroscience Institute,
    Princeton University, 
    Princeton, NJ 08540, USA%
    }
\author{David J.\ Schwab}\equalcontrib
    \affiliation{%
    Initiative for the Theoretical Sciences,
    The Graduate Center, CUNY, 
    New York, NY 10016, USA%
    }%
\author{Vudtiwat Ngampruetikorn}\equalcontrib
    \affiliation{%
    School of Physics, University of Sydney, 
    Sydney, NSW 2006, Australia%
    }
\begin{abstract}
Lossy data transformations by definition lose information. Yet, in modern machine learning, methods like data pruning and lossy data augmentation can help improve generalization performance.
We study this paradox using a solvable model of high-dimensional, ridge-regularized linear regression under \emph{data coarse graining}. Inspired by the renormalization group in statistical physics, we analyze coarse-graining schemes that systematically discard features based on their relevance to the learning task. 
Our results reveal a nonmonotonic dependence of the prediction risk on the degree of coarse graining. 
A \emph{high-pass} scheme---which filters out less relevant, lower-signal features---can help models generalize better. By contrast, a \emph{low-pass} scheme that integrates out more relevant, higher-signal features is purely detrimental. 
Crucially, using optimal regularization, we demonstrate that this nonmonotonicity is a distinct effect of data coarse graining and not an artifact of double descent.
Our framework offers a clear, analytical explanation for why careful data augmentation works: it strips away less relevant degrees of freedom and isolates more predictive signals. 
Our results highlight a complex, nonmonotonic risk landscape shaped by the structure of the data, and illustrate how ideas from statistical physics provide a principled lens for understanding modern machine learning phenomena. 
\end{abstract}
\maketitle

Information theory dictates that any irreversible modification of data can only decrease its information content~\cite{cover:06}; therefore, one might expect that it would not improve learning performance. Yet, modern machine learning abounds with counterexamples. One of the most intriguing phenomena is `double descent' which illustrates that \emph{decreasing} the size of training data can enhance generalization performance~\cite{belkin:19,nakkiran:20}. Other examples include techniques like data pruning
and data augmentation techniques (see, e.g., Refs~\cite{cubuk:19,sorscher:22}). Understanding how these phenomena emerge and how they may be leveraged in learning is of fundamental and practical importance.

Of course, such counterexamples do not disprove information theory. Typical learning algorithms do not utilize every available bit in the data. Indeed, not all bits are created equal. Some are relevant, others less so. Ideally, we want to learn the relevant bits and disregard the irrelevant ones. But, these bits are entangled: a complete dissociation between relevant and irrelevant bits is generally impossible~\cite{tishby:99}. Good algorithms are better at extracting more of the relevant bits and fewer of irrelevant ones~\cite{ngampruetikorn:22}. Additionally, practical constraints, such as computational costs and model expressivity and inductive biases, mean that some bits are harder to extract~\cite{xu:20}. Learning algorithms tend to rely on the bits that are easy to learn, regardless of their relevance. This tendency can lead to too much reliance on less relevant (but easier-to-learn) bits, resulting in suboptimal learning performance~\cite{dubios:20}. A question arises as to how we can encourage learning algorithms to focus on more relevant bits in the data.

We can explicitly penalize irrelevant information while maximizing relevant information during learning. This approach is equivalent to the information bottleneck method~\cite{tishby:99}, which has proved a powerful framework for preventing overfitting in deep learning~\cite{alemi:17,dubios:20}. However, estimating information-theoretic quantities from finite data is notoriously difficult. And, while variational bounds provide tractable estimates for information~\cite{poole:19}, they are generally loose and may not capture the salient behavior of information~\cite{tschannen:20}.

Here, we focus on a complementary approach which seeks to reduce irrelevant information in the data \emph{prior to} learning. While we cannot discard only irrelevant bits, in many interesting settings it is possible to identify and thus remove some of the less relevant bits. 

Indeed, this \emph{coarse graining} procedure forms the conceptual backbone of \emph{renormalization group}---a profoundly successful statistical physics framework that allows systematic investigations of collective effects in systems with many fluctuating components~\cite{goldenfeld:92,cardy:96}. In spatially extended systems, insights from physics suggest that long-wavelength fluctuations control the observable collective behavior and are therefore relevant. In this case, we can integrate out short-wavelength fluctuations without losing much information about the collective effects. The essence of this method is not to describe the systems with ever increasing accuracy but to reduce the full spectrum of complexity to a discrete set of universal behavior. This idea applies to not only physical systems but also a variety of biological systems from neural activity~\cite{meshulam:19,morales:23} to animal flocking~\cite{cavagna:19,cavagna:23}. Perhaps, we can exploit the same idea to tame the complexity of high-dimensional data~\cite{bradde:17}.

Although it is generally unclear what exact features in the data are relevant, we are not entirely clueless. For example, objects in images are often characterized by their shapes rather than the background. So, a good model for object classification should not rely on the bits that describe the background even though they may be correlated to label variables in specific datasets~\cite{ilyas:19,xiao:21}. It seems possible that removing these bits and forcing the model to make predictions based on more robust features could improve generalization performance. 
Indeed, over-specialization on dataset-specific features can be detrimental to out-of-distribution generalization~\cite{nguyen:24}.

To make this intuition more precise, we investigate the effects of data coarse graining on generalization performance. We consider a solvable model of ridge-regularized linear regression in the high-dimensional limit. We develop a theory that allows for a detailed study of several coarse graining schemes. Using both analytical results and simulations, we show that the schemes that integrate out weakly relevant features can improve generalization performance. Finally, we compare and contrast our results with standard double descent.

\begin{figure}
\centering
\includegraphics{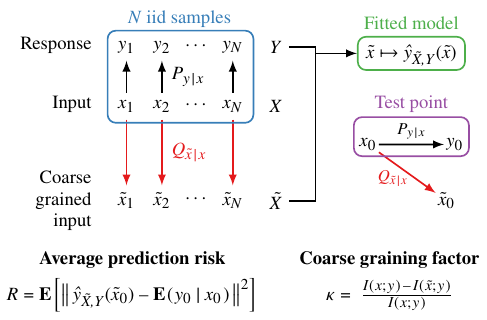}
\caption{\label{fig:cartoon}%
Schematic of the data coarse-graining procedure. We consider a supervised learning setup where the goal is to predict a response $y$ from an input feature vector $x$. Our framework introduces a coarse-graining step, a fixed channel $Q_{\tilde{x}|x}$, that transforms the original features $x$ into a new representation $\tilde{x}$ prior to learning. A model $\hat{y}(\cdot)$ is trained on a dataset of coarse-grained feature-response pairs. At test time, a new input $x_0$ is passed through the same channel to produce $\tilde{x}_0$ for prediction. Performance is measured by the \emph{average prediction risk,} which compares the model's output to the true conditional expectation, $\E(y_0|x_0)$. The information removed by the channel is quantified by the \emph{coarse-graining factor,} $\kappa$. This setup allows us to systematically study how the design of the channel $Q$ affects the final prediction risk.%
}
\end{figure}

Figure~\ref{fig:cartoon} provides the overview of our investigation. We consider a problem of learning a function $\hat y(\cdot)$ that predicts a response variable $y$ for a given feature (input) variable $x$. This function is chosen according to some learning algorithm $\mathcal L_{\hat y(\cdot)|\mathcal D}$ which takes as input, training data $\mathcal D$ of feature-response pairs, $(x_i,y_i)$ for $i=1,2,\dots,N$. We coarse grain the observed feature variable \emph{before} learning. That is, we replace every $x$ in the training data by their representation $\tilde x$, defined by some fixed channel $Q_{\tilde x|x}$, and as a result, the learning algorithm returns a model $\hat y(\cdot)$ that yields a prediction of a response $y$ for a transformed feature $\tilde x$. The data processing inequality means that this transformation can only decrease the information available for prediction---i.e., $I(\tilde x;y)\le I(x;y)$. However, this condition needs not imply poorer prediction performance as learning algorithms quite generally do not operate at information-theoretic limits. In fact, the inductive bias of model classes and fitting methods means that discarding part of the observed feature $x$---even if it decreases information about $y$---can result in better prediction performance.

To illustrate this phenomenon, we specialize to a simple setting of ridge-regularized linear regression. We assume that the feature variable $x\in\mathbb R^P$ is a Gaussian vector in $P$ dimensions with zero mean and a covariance matrix $\Sigma$, 
\begin{equation}\label{eq:x}
    x \sim \mathcal N(0,\Sigma). 
\end{equation}
The response $y\in\mathbb R$ to a feature $x$ is a scalar Gaussian variable with a mean that is linear in $x$ and variance $\sigma^2$,
\begin{equation}\label{eq:y|x}
    y\mid x \sim \mathcal N(\beta\tran x,\sigma^2).
\end{equation}
Here $\beta\in\mathbb R^P$ is the coefficient vector. For this generative model, the information between a feature vector and response reads
\begin{equation}\label{eq:I(x;y)}
    I(x;y) = \tfrac{1}{2}\ln (1+\snr),
\end{equation}
where $\snr = \beta\tran\Sigma\beta/\sigma^2$ is the signal to noise ratio. 

In addition, we consider a class of feature transformations, defined by a (noisy) projection, 
\begin{equation}\label{eq:z|x}
    \tilde x\mid x \sim \mathcal N(\Pi x,\omega^2I_P),
\end{equation}
where $\Pi\in\mathbb R^{P\times P}$ denotes a projection matrix, $I_P$ the identity matrix in $P$ dimensions and $\omega^2$ the projection noise variance. We note that since we define $\Pi$ to be a square matrix, the dimension of the representation $\tilde x$ is the same as that of $x$, i.e., $\tilde x\in\mathbb R^P$. In our setting, the learning algorithms have access to the representation $\tilde x$ and response $y$ but not the original feature vector $x$. Integrating out $x$ from Eqs~\eqref{eq:x},\,\eqref{eq:y|x}\,\&\,\eqref{eq:z|x} yields
\begin{equation}\label{eq:zy_gen}
    \tilde x \sim \mathcal N(0,\tilde\Sigma),\quad
    y\mid \tilde x \sim \mathcal N(\tilde\beta\tran \tilde x,\tilde\sigma^2),
\end{equation}
where $\tilde \Sigma = \Pi\Sigma\Pi\tran+\omega^2I_P$, $\tilde\beta=\tilde\Sigma^{-1}\Pi\Sigma\beta$ and $\tilde\sigma^2=\sigma^2+\beta\tran\Sigma\beta-\tilde\beta\tran\tilde\Sigma\tilde\beta$. The information about the response $y$ decreases as a result of this coarse graining transformation, 
\begin{equation}\label{eq:I(tildex;y)}
    I(\tilde x;y) = I(x;y) 
    - \tfrac{1}{2}\ln \big(1+(1-\eta)\snr\big),
\end{equation}
where \[\eta=\tilde\beta\tran\tilde\Sigma\tilde\beta/\beta\tran\Sigma\beta\] is the coarse-grained signal strength as a fraction of the original.
We use this information to define the \emph{coarse-graining factor},
\begin{equation}\label{eq:kappa}
    \kappa\equiv \frac{I(x;y)-I(\tilde x;y)}{I(x;y)} 
    =
    \frac{\ln(1+(1-\eta)\snr)}{\ln (1+\snr)}.
\end{equation}
We see that by definition $\kappa\in[0,1]$: $\kappa=0$ corresponds to no coarse graining and $\kappa=1$ to integrating out every bit that is predictive of the response $y$.

Linear regression assumes that the predictor is a linear map between a (transformed) feature $\tilde x$ and predicted response $\hat y$, 
\begin{equation}\label{eq:y-hat}
    \hat y(\tilde x) = \alpha\tran \tilde x,
\end{equation}
where $\alpha\in\mathbb R^P$ is a vector of model parameters. Given training data $\{(\tilde x_1,y_1),\dots,(\tilde x_N,y_N)\}$, we chose the parameters that minimize the ridge-regularized mean squared error, $\frac{1}{N}\sum_{i=1}^N(y_i-\hat y(\tilde x_i))^2+\lambda\|\alpha\|^2$ for $\lambda>0$, yielding 
\begin{equation}\label{eq:ridge-alpha}
    \alpha(\tilde X,Y) = (\tilde X\tilde X\tran+\lambda N I_P)^{-1} \tilde XY,
\end{equation}
where $Y\!=\!(y_1,y_2,\dots,y_N)\tran\!\in\!\mathbb{R}^N$ and $\tilde X\!=\!(\tilde x_1,\tilde x_2,\dots,\tilde x_N)\!\in\!\mathbb R^{P\times N}$. For clarity, we summarize the key parameters that define our model and analysis in Table~\ref{tab:variables}.

We measure the prediction performance by the expected risk
\begin{equation}\label{eq:risk}
    R = \E\nolimits_{\tilde X,Y,x,\tilde x} [\{\hat y(\tilde x;\tilde X,Y) - \E\nolimits_y(y \mid x)\}^2],
\end{equation}
where we make explicit the dependence of the predictor $\hat y(\cdot)$ on training data $(\tilde X,Y)$. Here the target of estimation is $\E_y(y\,|\,x)$ since we have access to the original feature vector $x$, even though it is not directly used in learning and predictions. Adding and subtracting $\E\nolimits_y(y \,|\,\tilde x)$ inside the curly brackets yields (see Ref~\cite[\S5]{hastie:22})
\begin{subequations}
\begin{align}
    R ={}& \E\nolimits_{\tilde X,Y,z} [\{\hat y(\tilde x;\tilde X,Y) - \E\nolimits_y(y \mid \tilde x)\}^2] \\
        &+ \underbrace{\E\nolimits_{x,\tilde x} [\{\E\nolimits_y(y \mid x) - \E\nolimits_y(y \mid \tilde x)\}^2]}_{=\mathcal M\;\;\text{(misspecification bias)}}.
        \label{eq:M}
\end{align}
\end{subequations}
The first term on the \rhs\ is the same as Eq~\eqref{eq:risk} but with $\E_y(y\,|\,\tilde x)$ as the estimation target. It is equivalent to the risk of predicting $y$ from $\tilde x$ without access to $x$. Adding and subtracting $\E\nolimits_Y(\hat y(\tilde x;\tilde X,Y) \,|\, \tilde X,\tilde x)$ inside the curly brackets of this term leads to the standard bias-variance decomposition
\begin{equation}\label{eq:R}
    R = \E\nolimits_{\tilde X} [\mathcal B_{\tilde X}] + \E\nolimits_{\tilde X} [\mathcal V_{\tilde X}] + \mathcal M,
\end{equation}
where $\mathcal B_{\tilde X}$ and $\mathcal V_{\tilde X}$ denote the $\tilde X$-dependent bias and variance, defined by
\begin{align}
    \label{eq:B_tildeX_E}
    \mathcal B_{\tilde X} &= \E_{Y,\tilde x} [\{\E\nolimits_Y(\hat y(\tilde x;\tilde X,Y) \,|\, \tilde X,\tilde x) - \E\nolimits_y(y \mid \tilde x)\}^2\mid \tilde X]
    \\
    \label{eq:V_tildeX_E}
    \mathcal V_{\tilde X} &= \E_{Y,\tilde x} [\{\hat y(\tilde x;\tilde X,Y) - \E\nolimits_Y(\hat y(\tilde x;\tilde X,Y) \mid \tilde X,\tilde x)\}^2 \mid \tilde X].
\end{align}
The last term in Eq~\eqref{eq:R} is known as the \emph{misspecification bias}~\cite{hastie:22}, which is the irreducible estimation bias from using the coarse grained feature instead of the original one. For the generative model in Eqs~\eqref{eq:x},\,\eqref{eq:y|x},\,\eqref{eq:z|x}\,\&\,\eqref{eq:zy_gen}, this term is given by
\begin{equation}\label{eq:mathcalM}
    \mathcal M 
    = 
    \beta\tran \Sigma\beta - \tilde\beta\tran \tilde\Sigma\tilde\beta
    =
    \sigma^2(1-\eta)\snr.
\end{equation}
We see that this bias is equal to the reduction in signal strength as we transform the feature $x$ into its representation $\tilde x$.

For linear predictors [Eq~\eqref{eq:y-hat}], Eqs~\eqref{eq:B_tildeX_E}\,\&\,\eqref{eq:V_tildeX_E} reduces to
\begin{align}
    \mathcal B_{\tilde X} &= \big\|\tilde\Sigma^{1/2}(\mathbf E_{Y}(\alpha(\tilde X,Y)\mid \tilde X)-\tilde\beta)\big\|^2
    \\
    \mathcal V_{\tilde X} &= \tr\left(\tilde\Sigma\Cov\nolimits_Y( \alpha(\tilde X,Y) \mid \tilde X)\right),
\end{align}
where $\tilde\Sigma$ and $\tilde\beta$ are defined in Eq~\eqref{eq:zy_gen} and $\Cov(a)=\E[(a-\E[a])(a-\E[a])\tran]$ is the covariance matrix of a random vector $a$. Substituting the ridge estimator for $\alpha$ [Eq \eqref{eq:ridge-alpha}] into the above equations gives
\begin{align}
    \mathcal B_{\tilde X}
    &= 
    \tilde\beta\tran
    \frac{\lambda}{\Psi+\lambda I_P}
    \tilde\Sigma
    \frac{\lambda}{\Psi+\lambda I_P}
    \tilde\beta
    \label{eq:B_tildeX}
    \\
    \mathcal V_{\tilde X}
    &= \tilde\sigma^2 
    \frac{1}{N}
    \tr\left(\tilde\Sigma\frac{\Psi}{(\Psi+\lambda I_P)^2}\right),
    \label{eq:V_tildeX}
\end{align}
where $\Psi = \frac{1}{N}\tilde X\tilde X\tran$ is the empirical covariance matrix of the coarse grained feature vector.

We now let $P, N\to\infty$ while fixing the ratio $\gamma=P/N\in(0,\infty)$. In this asymptotic limit, the above bias and variance terms become deterministic~\cite{wu:20,richards:21,hastie:22}
\begin{align}
    \label{eq:mathcalB}
    \mathcal B_{\tilde X}
    \to\mathcal B &= 
    \|\tilde\beta\|^2\lambda^2\nu'(-\lambda)
    \frac{
    \int\!\!dG^{\tilde\Sigma,\tilde \beta}(s)
    \frac{s}{(s + \lambda(1+\gamma \nu(-\lambda)))^2}
    }{
    \int\!\!dF^{\tilde\Sigma}(s)
    \frac{s}{(s + \lambda(1+\gamma \nu(-\lambda)))^2}
    }
    \\
    \label{eq:mathcalV}
    \mathcal V_{\tilde X}
    \to \mathcal V &= 
    \sigma^2(1+(1-\eta)\snr) \gamma\left(
     \nu(-\lambda)
    -
    \lambda\nu'(-\lambda)
    \right),
\end{align}
where $F^{A}$ denotes the cumulative spectral distribution of a real symmetric matrix $A$ and $G^{A,b}$ the same distribution but with the weight of each eigenmode replaced by $(b\tran u_i)^2/\|b\|^2$ where $u_i$ is the eigenvector of that mode and $b$ is some real vector. Here the function $\nu(z)$ is the unique upper half-plane solution of the self-consistent equation~\footnote{Note that $\nu(z)$ is related to $m(z)$ in Refs~\cite{hastie:22,richards:21} via $m(z)=\frac{\nu(z)}{z(1+\gamma\nu(z))}-\frac{1}{z}$.},
\begin{equation}\label{eq:nu_z}
    \nu(z) 
    = 
    \int\!\!dF^{\tilde\Sigma}(s) \frac{s}{\frac{1}{1+\gamma\nu(z)}s -z}.
\end{equation}
Equations \eqref{eq:mathcalM},\,\eqref{eq:mathcalB}\,\&\,\eqref{eq:mathcalV} completely characterize the average prediction risk in the asymptotic limit.

To isolate the effects of data coarse graining, we turn to the simple case of isotropic features $\Sigma=I_P$ and diagonal coarse graining projection $\Pi=\diag(\pi_1,\dots,\pi_P)$ with $\pi_i\in\{0,1\}$~\footnote{Recall that the an eigenvalue of a projection matrix is either one or zero.}. We specialize further to noiseless coarse graining, taking the limit $\omega^2\to0^+$. In this case, the coarse-grained covariance matrix and coefficient vector become $\tilde\Sigma=\Pi$ and $\tilde\beta=\Pi\beta$. The bias term, Eq~\eqref{eq:mathcalB}, reduces to 
\begin{equation}
    \label{eq:mathcalB_iso}
    \mathcal B = 
    \sigma^2\eta\,\snr\,\lambda^2\nu'(-\lambda)/q,
\end{equation}
where \[q\equiv\tfrac{1}{P}\tr \Pi\] is the fraction of nonzero eigenvalues of $\Pi$. Solving Eq~\eqref{eq:nu_z} yields closed-form expressions for $\nu(z)$ and its derivative $\nu'(z)$,
\begin{equation}
\label{eq:nu_iso}
\begin{aligned}
\nu(z) &= 
    \frac{1}{2\gamma z}
    \Big[1-q\gamma-z-\sqrt{(1-q\gamma-z)^2-4q\gamma z}\Big]
    \\
\nu'(z) &= 
    \frac{q(1+\gamma\nu(z))^2}{[1-z(1+\gamma\nu(z))]^2-q\gamma}.
\end{aligned}
\end{equation}
We obtain the thermodynamic-limit prediction risk by substituting the above into Eqs~\eqref{eq:mathcalV}\,\&\,\eqref{eq:mathcalB_iso} and combining the resulting variance and bias terms with the misspecification bias, Eq~\eqref{eq:mathcalM}.

\begin{figure*}
\centering
\includegraphics[width=\linewidth]{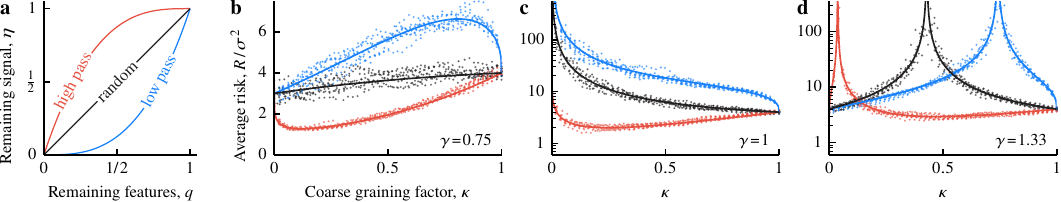}
\caption{\label{fig:risks_min-norm}%
Data coarse graining can improve generalization performance. 
\textbf{a} Coarse graining schemes are defined by the relationship between the remaining signal strength, $\eta\!=\!\tilde\beta\tran\tilde\Sigma\tilde\beta/\beta\tran\Sigma\beta$, and the remaining number of features, $q\!=\!\frac{1}{P}\tr\Pi$. Here the entries of the original coefficient vector $\beta$ are \iid\ random variables, $\beta_i\!\sim\!\mathrm{Uniform}(-1,1)$. The high-pass coarse graining scheme (red curve) integrates out low-signal features first; as a result, coarse graining away the first few features (i.e., decreasing $q$ from one) leaves the signal strength $\eta$ relatively unchanged. The low-pass scheme (blue curve) discard high-signal features, thus resulting in a sharp drop in $\eta$ near $q\!=\!1$. Eliminating features at random does not alter the distribution of the signal and the average signal strength per feature is constant, i.e., $\eta\!=\!q$ (black line). 
\textbf{b-d} Prediction risk \emph{vs} coarse graining factor [see Eq~\eqref{eq:kappa} for definition] for min-norm regression [Eq~\eqref{eq:minnormrisk}] at different parameter-to-sample ratios $\gamma\!=\!P/N$ (see labels) under coarse graining schemes displayed in Panel a (same legend). 
We see that the average risk exhibits nonmonotonic dependence on the coarse graining factor. For the high-pass scheme, the risk is smallest at intermediate coarse graining. For $\gamma\!>\!1$, coarse graining can induce a transition from over to underparametrized regimes at which the risk diverges (Panels c\,\&\,d). 
The theoretical predictions (solid curves) agree well with the simulations (symbols). 
In Panels b-d, we set $\snr\!=\!4$. We simulate ten realizations of $(\beta,X,Y)$ with $N\!=\!1200$ and $P\!=\!900,1200,1600$ (Panels b-d) and each symbol shows the empirical prediction risk averaged over 1000 test points.
}
\end{figure*}

It is instructive to first consider the ridgeless limit $\lambda\to0^+$, which admits a simpler analytical solution for prediction risk. Expanding $\nu(z)$ around $z=0$ yields
\begin{equation}
    \nu(z) = \frac{1}{\gamma}\times\left\{
    \begin{array}{ll}
    \frac{q\gamma}{1-q\gamma} + O(z),
        &q\gamma<1\\
    -\frac{q\gamma-1}{z}
    + \frac{1}{q\gamma-1} + O(z),     
        &q\gamma>1  \\
    \end{array}\right.
\end{equation}
Substituting this solution into Eqs~\eqref{eq:mathcalB}\,\&\,\eqref{eq:mathcalV} gives (c.f.~Ref~\cite{hastie:22})
\begin{align}
    \mathcal B &= \sigma^2\eta\,\snr\times
    \left\{
    \begin{array}{ll}
    O(\lambda^2),   &q\gamma<1\\
    1-\frac{1}{q\gamma}+O(\lambda),   &q\gamma>1\\
    \end{array}\right.
    \\
    \mathcal V &= 
    \sigma^2(1+(1-\eta)\snr) \times
    \left\{
    \begin{array}{ll}
    \frac{q\gamma}{1-q\gamma}+O(\lambda),   &q\gamma<1\\
    \frac{1}{q\gamma-1}+O(\lambda),   &q\gamma>1\\
    \end{array}\right..
\end{align}
Combining these contributions with the misspecification bias [Eq~\eqref{eq:mathcalM}], we obtain the average risk 
\begin{align}\label{eq:minnormrisk}
    \frac{R}{\sigma^2} &=
    \left\{
    \begin{array}{ll}
    \frac{(1-\eta)\snr\,+\,q\gamma}{1-q\gamma},   
    &q\gamma<1\\
    \frac{(1-\eta)\snr\,+\,1/q\gamma}{1-1/q\gamma}
    +\eta\,\snr (1-1/q\gamma),   
    &q\gamma>1\\
    \end{array}\right.,
\end{align}
where we drop the terms of order $O(\lambda)$ and smaller. This expression is identical to that in Ref~\cite[\S3]{hastie:22} when $\eta=q=1$ (i.e., no coarse graining). In our setting, the effective signal strength $\eta$ and number of features $q$ decrease with the degree of data coarse graining.

\begin{table}[t!]
\centering
\caption{Key parameters and variables in our analysis.}
\label{tab:variables}
\begin{tabular}{@{}c@{\quad}p{6.7cm}@{}}
\toprule
\textbf{Symbol} & \textbf{Description} \\ \midrule
$P$, $N$ & Number of features and samples, respectively. \\
$\gamma$ & Parameter-to-sample ratio ($P/N$). \\
\addlinespace
$q$ & Fraction of features remaining after coarse graining. \\
$\eta$ & Fraction of signal remaining after coarse graining.\\ 
$\kappa$ & Fraction of predictive information integrated out.\\
\addlinespace
$\lambda$ & Ridge regularization strength. \\
$\snr$ & Signal-to-noise ratio. \\ \bottomrule
\end{tabular}
\end{table}

Indeed the relationship between the remaining signal strength $\eta$ and the effective number of remaining features $q$ describes the coarse-graining procedure. In Fig.~\ref{fig:risks_min-norm}a, we visualize three distinct schemes. The \emph{high-pass} scheme (red curve) models a procedure that integrates out low-signal features first. This preferential removal of less relevant features means the signal strength $\eta$ decreases more slowly than the feature count $q$ at the coarse graining onset (near $\eta=q=1$). 
On the other hand, the \emph{low-pass} scheme (blue curve) represents the case where more relevant features are discarded first, resulting in a rapid loss of predictive information near $\eta=q=1$ where $\eta$ decays faster than $q$. 
Finally, the \emph{random-pass} scheme (black line) provides a neutral baseline. By removing features at random regardless of their signals, the remaining signal strength $\eta$ decreases linearly with the feature count $q$, yielding $\eta=q$ over the entire course of the procedure~\footnote{While we analyze these archetypal schemes, the set of physically achievable trajectories is constrained by the data's underlying signal distribution. If signal strength were distributed uniformly across all features, any feature removal procedure would trace the random-pass path. The high and low-signal-pass schemes are realizable only when the signal is nonuniform, allowing for preferential selection.}.

Figure~\ref{fig:risks_min-norm} depicts the average prediction risk of ridge regression as a function of the coarse-graining factor $\kappa$, illustrating the effects of coarse-graining schemes (Fig~\ref{fig:risks_min-norm}a) in three different parameter-to-sample regimes (panels b-d). We see that prediction risk depends on the coarse-graining schemes, the degree of coarse-graining $\kappa$ and the parameter-to-sample ratio $\gamma=P/N$. In all cases, we observe good agreement between theoretical predictions [Eq~\eqref{eq:minnormrisk}] (solid lines) and direct simulations (circles).

In the under-parameterized regime, also known as the classical or data-abundant setting, where the sample size $N$ is greater than the number of features $P$ (i.e., $\gamma<1$), we see that the impact of coarse graining depends strongly on the scheme (Fig~\ref{fig:risks_min-norm}b). The high-pass scheme, which by design preserves the most relevant features, leads to a nonmonotonic risk curve with a distinct minimum at an intermediate $\kappa$. That is, integrating out less relevant features can improve generalization. 
But this phenomenon is not universal. The low-pass scheme, which discards high-signal features first, also yields a nonmonotonic risk but with a maximum (instead of a minimum) at an intermediate $\kappa$. 
As expected, the random-pass scheme provides a smooth interpolating behavior between these two extremes.

In the over-parameterized regime ($\gamma\ge1$), we see that the high-pass scheme also decreases the risk with a minimum at an intermediate value of $\kappa$ (see Fig~\ref{fig:risks_min-norm}c\,and\,d). However, the risk landscape is now dominated by a singularity at the interpolation threshold. As a result, all coarse-graining schemes exhibit a risk divergence, as reducing the effective feature count $q$ eventually drives the system across this threshold, from over to under-parameterized. This singularity marks the interpolation transition---at which the model minimizes the loss by passing through all of the training data---and is related to the double descent phenomena (see, e.g., Refs~\cite{hastie:22,wu:20,richards:21}). We note that while a true divergence is an artifact of vanishing regularization $\lambda\to0^+$, the qualitative behavior of the risk is unchanged away from such limits.

\begin{figure}
\centering
\includegraphics{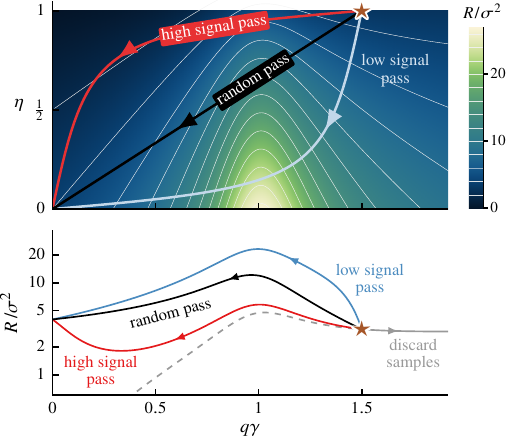}
\caption{\label{fig:risk-heatmap}%
Risk landscape underpins the generalization performance of coarse-graining schemes.
Prediction risk landscape (color map) as a function of the effective parameter-to-sample ratio, $q\gamma$, and the remaining signal strength, $\eta$. 
The paths show three coarse-graining schemes from Fig~\ref{fig:risks_min-norm}a (see labels) for $\gamma\!=\!1.5$. Each path begins at $q\!=\!\eta\!=\!1$ (star) and moves left as features are removed. 
For comparison, we also include a path for standard double descent (gray curve), where all features are preserved ($q\!=\!\eta\!=\!1$) while the number of samples, and thus $\gamma$, is varied, tracing a horizontal path on the landscape, with the dashed portion to the left of the star symbol accessible only by adding more data.
We see that the landscape is dominated by a ridge of high risk centered around the interpolation threshold at $q\gamma\!=\!1$. The path corresponding to the high-pass scheme traverses close to the top left corner, where the remaining signal is high ($\eta\!\approx\!1$) and the data becomes abundant compared to the effective number of features ($N\!\gg\!qP$).
Here we compute prediction risk from Eqs \eqref{eq:mathcalM},\,\eqref{eq:mathcalV},\,\eqref{eq:mathcalB_iso}\,\&\,\eqref{eq:nu_iso} with $\snr\!=\!4$, $\omega^2\!\to\!0^+$ and $\lambda\!=\!0.01$.%
}
\end{figure}

In these examples, each coarse-graining scheme corresponds to a path across the risk landscape parametrized by the effective parameter-to-sample ratio $q\gamma$ and remaining signal strength $\eta$, see Fig~\ref{fig:risk-heatmap}. This risk landscape is dominated by a maximum at interpolation threshold $q\gamma=1$. When $\gamma>1$, all coarse-graining paths begin in the over-parameterized region ($q\gamma>1$) and cross this ridge to the under-parameterized side, resulting in the distinctive double descent peak (see also Fig~\ref{fig:risks_min-norm}c and d). For comparison, we plot the risk for the case where we discard samples while preserving all available features (i.e., increase $\gamma$ while fixing $q=\eta=1$). In this case, the system follows a horizontal line across this heatmap, exhibiting standard double-descent behavior. Note that for ease of visualization, we set $\lambda=0.01$; the ridge at $q\gamma=1$ diverges as $\lambda\to0^+$.

\begin{figure}
\centering
\includegraphics{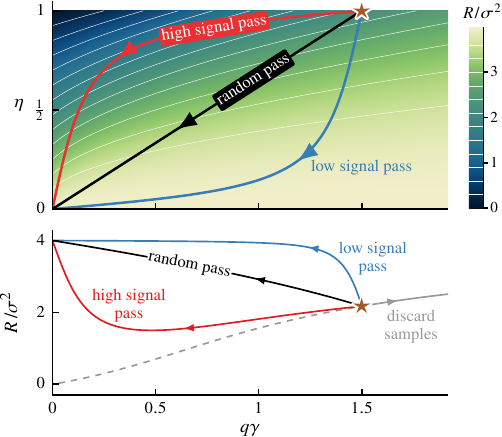}
\caption{\label{fig:risk-heatmap-optimal}%
Optimal regularization isolates risk nonmonotonicity unique to coarse graining. 
Prediction risk landscape for optimally tuned ridge regression [Eq~\eqref{eq:risk-optimal}]. The axes and overlaid paths are the same as in Fig~\ref{fig:risk-heatmap}. Optimal ridge regularization [Eq~\eqref{eq:lambda-optimal}] completely suppresses the double-descent structure at $q\gamma=1$ (cf.~Fig~\ref{fig:risk-heatmap}). 
Here, the risk behavior of the high-pass scheme remains nonmonotonic, exhibiting a minimum risk at an intermediate level of coarse graining. Parameters (except for $\lambda$) are the same as in Fig~\ref{fig:risk-heatmap}.%
}
\end{figure}

The divergence at the interpolation threshold can obscure the more subtle effects of coarse graining. To better isolate the impact of data representation, we now analyze \emph{optimally tuned ridge regression,} which completely suppresses double descent in prediction risk~\cite{dobriban:18,nakkiran:21,hastie:22}. Minimizing prediction risk---see Eqs \eqref{eq:mathcalM},\,\eqref{eq:mathcalV},\,\eqref{eq:mathcalB_iso}\,\&\,\eqref{eq:nu_iso}---yields the optimal regularization strength (for each set of \snr, $q\gamma$ and $\eta$)
\begin{equation}\label{eq:lambda-optimal}
    \lambda^*=\frac{q\gamma(1+(1-\eta)\,\snr)}{\eta\,\snr},
\end{equation}
and the optimally tuned prediction risk
\begin{align}
\frac{R^*}{\sigma^2}={}&\snr - \frac{1}{2}
\Bigg\{
1+\snr+\frac{\eta\,\snr}{q\gamma}
\nonumber\\&\qquad
-\sqrt{
\Big(1+\snr+\frac{\eta\,\snr}{q\gamma}\Big)^2-4q\gamma \Big(\frac{\eta\,\snr}{q\gamma}\Big)^2
}
\Bigg\}.
\label{eq:risk-optimal}
\end{align}
In the infinite-data limit ($\gamma\to0^+$), the optimally tuned risk $R^*$ approaches the irreducible error floor set by the misspecification bias, $\sigma^2(1-\eta)\snr$. In contrast, when all signal is removed ($\eta=0$), the risk is maximum with $R^*=\sigma^2\,\snr$.

Figure~\ref{fig:risk-heatmap-optimal} depicts the optimally tuned prediction risk landscape [Eq~\eqref{eq:risk-optimal}]. We see that the double-descent ridge at the interpolation threshold ($q\gamma=1$) is entirely absent (cf.~Fig~\ref{fig:risk-heatmap}). The risk landscape is now smooth, enabling a clear assessment of the impact of each coarse-graining scheme and revealing that the double-descent peak is not required for risk nonmonotonicity. We depict three coarse-graining schemes (same as in Fig~\ref{fig:risk-heatmap}). The low-pass and random-pass schemes now increase the prediction risk monotonically. In contrast, the high-pass scheme still results in minimum risk at an intermediate coarse-graining level.

In this work, we investigate the paradoxical phenomenon where reducing the information content of data can enhance generalization. Our analysis of a solvable, high-dimensional regression model reveals that the effect of such coarse graining depends on the procedure. A high-pass scheme that preferentially removes low-signal features can lead to a nonmonotonic risk curve with a minimum at an intermediate level of coarse graining. In contrast, a low-pass scheme that removes high-signal features is consistently detrimental. By analyzing the system under optimal regularization, we demonstrated that this nonmonotonicity is an intrinsic effect of coarse graining, fundamentally distinct from the double-descent phenomenon, which induces divergent prediction risk at the interpolation threshold.

These findings can be understood in the spirit of the renormalization group (RG). The high-pass coarse-graining scheme is analogous to an RG-like procedure that integrates out less relevant degrees of freedom---the low-signal features---thus improving the effective signal-to-noise ratio of the data. We can understand this mechanism through the decomposition of prediction risk. While any removal of signal necessarily increases the misspecification bias, this increase is minimal for a high-pass scheme. Our results show that this cost is more than compensated by a reduction in estimator variance, leading to a net improvement in overall prediction risk.

This theoretical framework helps contextualize the performance of common heuristics in machine learning. For example, techniques like data augmentation can be viewed as implicitly encouraging a model to become invariant to low-relevance details, conceptually similar to a high-pass coarse-graining scheme. Our work also clarifies the behavior of feature selection methods, suggesting that their success is not guaranteed by dimensionality reduction alone, but depends also on the criterion used for removal. A procedure that successfully identifies and discards irrelevant features, even on average, can improve performance, whereas those that discard relevant features can hurt generalization.

Our results complement recent findings on Principal Component Regression (PCR). \citet{gedon:24} showed that PCR, an unsupervised method, suppresses double descent by reducing the effective number of features, thus keeping the system away from the interpolation threshold. This result demonstrates how a variance-based feature removal can act as an effective regularizer. Our work, in contrast, investigates data coarse graining defined by the relevance of each feature to the prediction target, thus making it a supervised technique. We find that data coarse graining introduces a distinct form of nonmonotonic risk that persists even when double descent is suppressed via optimal regularization. Taken together, these results highlight the rich effects of feature removal. While PCR can regularize primarily by preventing interpolation, data coarse graining is underpinned by the tradeoff between signal fidelity and estimator variance.

Our work analyzes a solvable linear regression model. We assume isotropic features to ensure analytical tractability. This approach allowed us to isolate the mechanisms by which coarse graining affects generalization. While essential for clarity, these simplifying assumptions also define several promising directions for future research. It would be interesting to extend these principles to models with more complex data structures, such as anisotropic feature covariances, and to investigate whether these principles apply to non-linear settings, such as kernel methods or simple neural networks, where the interplay between data representation and model capacity is even more complex. Furthermore, developing a framework for more realistic, data-driven coarse-graining procedures would bridge the gap between our theoretical model and the successful heuristics of practical data augmentation. Ultimately, a deeper theoretical understanding of these processes can provide a foundation for developing principled data augmentation techniques, leading to more effective use of large datasets and complex models.

\begin{acknowledgments}
Alex Nguyen is supported by NIH grant RF1MH125318.
DJS was partially supported by a Simons Fellowship in the MMLS, a Sloan Fellowship, and the National Science Foundation, through the Center for the Physics of Biological Function (PHY-1734030). 
VN acknowledges research funds from the University of Sydney.
\end{acknowledgments}

\end{document}